\documentclass{llncs}

\usepackage[british]{babel}
\usepackage{amsfonts}
\usepackage{booktabs}
\usepackage{svn-multi}
\usepackage{url}
\usepackage{xspace}
\usepackage{graphicx}
\usepackage{tikz}
\usetikzlibrary{calc}
\usetikzlibrary{positioning}
\usetikzlibrary{automata}
\usetikzlibrary{shapes}
\usetikzlibrary{decorations.pathreplacing}
\usepackage{wrapfig}
\usepackage{times}
\usepackage{amsmath}

\usepackage[caption=false]{subfig}
\usepackage{listings}
\newcommand{\tv}[1]{\textcolor{magenta}{\ifmmode \text{[TV: #1]}\else [TV: #1] \fi}}
\newcommand{\ps}[1]{\textcolor{cyan}{\ifmmode \text{[VH: #1]}\else [VH: #1] \fi}}
\newcommand{\vm}[1]{\textcolor{green}{\ifmmode \text{[VM: #1]}\else [VM: #1] \fi}}
\newcommand{\mh}[1]{\textcolor{blue}{\ifmmode \text{[MH: #1]}\else [MH: #1] \fi}}

\newcommand{\dynobj}[2]{ao_{#1}^{#2}}
\newcommand{\nullptr}{\mathtt{NULL}}

\makeatletter
\let\UrlSpecialsOld\UrlSpecials
\def\UrlSpecials{\UrlSpecialsOld\do\/{\Url@slash}\do\_{\Url@underscore}}%
\def\Url@slash{\@ifnextchar/{\kern-.11em\mathchar47\kern-.2em}%
    {\kern-.0em\mathchar47\kern-.08em\penalty\UrlBigBreakPenalty}}
\def\Url@underscore{\nfss@text{\leavevmode \kern.06em\vbox{\hrule\@width.3em}}}
\makeatother

\title{2LS: Heap Analysis and Memory Safety\vspace*{-1mm}}
\subtitle{(Competition Contribution)\thanks{The Czech authors were supported by
  the Czech Science Foundation project 17-12465S, H2020 ECSEL Aquas project,
  the IT4IXS: IT4Innovations Excellence in Science project (LQ1602), and the
  FIT BUT internal project FIT-S-17-4014.}\vspace*{-3mm}}

\author{
  Martin Hru\v{s}ka\inst{3} \and
  Viktor Mal\'{\i}k\inst{3} \and
  Peter Schrammel\thanks{Jury member: \textsf{p.schrammel@sussex.ac.uk}.}\inst{1,2} \and
  \mbox{Tom\'{a}\v{s} Vojnar}\inst{3}
}

\institute{
  ~\inst{1}Diffblue Ltd, Oxford, UK \hspace*{3mm}
  ~\inst{2}University of Sussex, Brighton, UK \\
  ~\inst{3}FIT BUT, IT4Innovations Centre of Excellence, CZ \hspace*{3mm}
}

\begin{document}

\maketitle

\vspace*{-4mm}\begin{abstract}
2LS is a framework for analysis of sequential C programs that can verify and
refute program assertions and termination.
The 2LS framework is built upon the CPROVER infrastructure and implements
template-based synthesis techniques, e.g.\ to find invariants and ranking
functions, and incremental loop unwinding techniques to find counterexamples and
$k$-induction proofs.
The main improvements in this year's version are the ability of 2LS to analyse
programs requiring combined reasoning about shape and content of dynamic data
structures, and an instrumentation for memory safety properties.
\end{abstract}

\section{Overview} \label{sec:introduction}

2LS is a static analysis and verification tool for sequential C
programs. At its core, it uses the $k$I$k$I algorithm
(\mbox{$k$-invariants} and $k$-induction)~\cite{BJKS15}, which
integrates bounded model checking, $k$-induction, and abstract
interpretation into a single, scalable framework. $k$I$k$I relies on
incremental SAT solving in order to find proofs and refutations of
assertions, as well as to perform termination analysis~\cite{CDK+17}.

This year's competition version introduces new product and power
domain combinations to support invariant inference for programs that
manipulate shape and content of dynamic data structures~\cite{MHSV18}.
Moreover, there is an improved encoding of memory safety properties.

\paragraph{\bf Architecture.} \label{sec:architecture}

The architecture of 2LS has been described in previous
competition contributions~\cite{SK16,MMSSVW18}.
In brief, 2LS is built upon the CPROVER
infrastructure~\cite{DBLP:conf/tacas/ClarkeKL04} and thus uses
\emph{GOTO programs} as the internal program representation.
The analysed program is
translated into an acyclic,
over-approximate single static assignment (SSA) form, in which loops
are cut at the edges returning to the loop head. Subsequently, 2LS
refines this over-approximation by computing inductive invariants in
various abstract domains represented by parametrised logical formulae,
so-called templates~\cite{BJKS15}. The competition version uses
the zones domain for numerical variables in combination with
our shape domain for pointer-typed variables and the symbolic paths
domain described below.
The SSA form is bit-blasted into a propositional formula and given to a
SAT solver. The $k$I$k$I algorithm then incrementally amends the formula
to perform loop unwindings and invariant inference based on
template-based synthesis~\cite{BJKS15}.

\section{New Features}

The major improvements for SV-COMP'20 are all related to analysis of
heap-{ma\-ni\-pu\-la\-ting} programs. We build on the shape domain
presented last year~\cite{MMSSVW18} and introduce
abstract domain combinations that allow us to analyse both shape and
content of dynamic data structures.
Furthermore, we present encoding of
assertions that are used for verifying memory safety properties.

\subsection{Combinations of Abstract Domains}

The new capability of 2LS to jointly analyse shape and content of dynamic
data structures takes advantage of the template-based synthesis
engine of 2LS. Invariants are computed in various abstract domains where each
domain has the form of a template
while relying on the
analysis engine to handle the domain combinators.

\paragraph{\bf Memory model}

In our memory model, we represent dynamically allocated objects by
so-called \emph{abstract dynamic objects}. Each such object is an
abstraction of a number of concrete dynamic objects allocated by the
same \textsf{malloc} call (i.e. at the same program location)~\cite{MHSV18}.

\paragraph{\bf Shape Domain}

For analysing the shape of the heap, we use an improved version of the
shape domain that we introduced last year~\cite{MMSSVW18}. The domain
over-approximates the \emph{points-to} relation between pointers and
symbolic addresses of memory objects in the analysed program: for each
pointer-typed variable and each pointer-typed field of an abstract
dynamic object $p$, we compute the set of all addresses that $p$ may
point to~\cite{MHSV18}.

\paragraph{\bf Template Polyhedra Domain}

For analysing numerical values, we use the template polyhedra abstract
domains, particularly the \emph{interval} and the \emph{zones}
domains~\cite{BJKS15}.

\paragraph{\bf Shape and Polyhedra Domain Combination}

Since both domains have the form of a~template formula, we
simply use them side-by-side in a product domain combination---the
resulting formula is a conjunction of the two template
formulae~\cite{MHSV18}.

\begin{wrapfigure}[8]{r}{17em}
  \centering
  \vspace{-5ex}
\begin{tikzpicture}[>=latex]
  \tikzstyle {field} = [draw, rectangle, minimum height = 1.2em, 
  minimum width = 4em,  node distance = 1.2em, font=\ttfamily\scriptsize];
  \tikzstyle {pointer} = [node distance = 4em, font=\ttfamily\scriptsize];
  \tikzstyle {nextptr} = [->]

  \node [field] (next1) {next};
  \node [field, below of = next1] (val1) {val$=3$};

  \node [field, right of = next1, node distance = 9.5em] (next2) {next};
  \node [field, below of = next2] (val2) {val$=10$};

  \node [pointer, right of = next2] (null) {$\nullptr$};

  \path [nextptr] (next1) edge +(1, 0);
  \path ($(next1)+(1.1,0)$) --  +(1,0) node [font=\Large, midway, sloped] {$\dots$};
  \path [nextptr] ($(next1)+(2,0)$) edge (next2);
  \path [nextptr] (next2) edge (null);

  \draw [decoration={brace, raise=12pt}, decorate] 
  ($(next1)+(-.5,0)$) -- node[above=14pt]{$\dynobj{1}{}$} ($(next2)+(.5,0)$);
\end{tikzpicture}
\caption{Unbounded singly-linked list abstracted by an abstract dynamic object $\dynobj{1}{}$.}
\label{fig:list}
\end{wrapfigure}
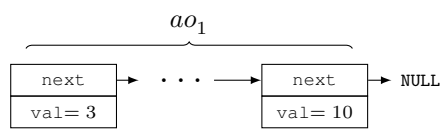
\noindent This combination allows 2LS to infer, e.g., invariants describing an
unbounded singly-linked list whose nodes contain values between 1 and 10. We
show an example of such a list in Figure~\ref{fig:list}.
Here, all list nodes are abstracted by a single abstract dynamic object
$\dynobj{1}{}$ (i.e. we assume that they are all allocated at the same program
location). The invariant inferred by 2LS for such a list might look as
follows:\vspace*{-2mm}
\begin{equation}
  (\dynobj{1}{}.next = \&\dynobj{1}{} \vee \dynobj{1}{}.next = \nullptr) \wedge
  \dynobj{1}{}.val \in [1, 10]
\end{equation}
The first disjunction describes the shape of the list---the \emph{next} field
of each node points to some node of the list or to $\nullptr$\footnote{Here,
$\dynobj{1}{}.f$ is an abstraction of the $f$ fields of all concrete objects
represented by $\dynobj{1}{}$. Analogously, $\&\dynobj{1}{}$ is an
abstraction of symbolic addresses of all represented objects.}. The second
part of the conjunct is then an invariant in the interval domain over all
values stored in the list---it expresses the fact that the value of each node
lies in the interval between 1 and 10.

\subsection{Symbolic Paths}

To improve precision of the analysis, we let 2LS compute different
invariants for different \emph{symbolic paths} taken by the analysed
program. We require a symbolic path to express which loops were
executed at least once. This allows us to distinguish situations when
an abstract dynamic object does not represent any really allocated
object and hence the invariant for such abstract dynamic object is not
valid~\cite{MHSV18}.
The symbolic path domain allows us to iteratively compute a set of
symbolic paths $p_1, \dots, p_n$ (represented by guard variables in
the SSA) with associated shape and data invariants $I_1, \dots, I_n$.
The aggregated invariant is then $p_1 \Rightarrow I_1 \wedge
\dots \wedge p_n \Rightarrow I_n$, which corresponds to a power
domain combination.

\subsection{Memory Safety}

To verify memory safety, appropriate assertions are inserted into the program.
We now describe the structure of these assertions for different types of memory
errors.

\paragraph{\bf Dereferencing/Freeing a $\nullptr$ Pointer}

To check for this kind of errors, we add an assertion $p \neq \nullptr$ to each
location where \textsf{*p} or \textsf{free(p)} occurs~\cite{MHSV18}. Since the
shape domain over-approximates the set of all addresses that $p$ may point to,
absence of such errors can be proven. If an error is found, we use BMC to
check its reachability.

\paragraph{\bf Dereferencing/Freeing a Freed Pointer}

Using a single abstract dynamic object to represent multiple concrete
objects poses problems when trying to determine if a particular
concrete object (within the abstract one) has already been freed or
not. To resolve this, for each abstract dynamic object $\dynobj{i}{}$,
we non-deterministically select a single concrete object represented
by $\dynobj{i}{}$ and materialize it as $\dynobj{i}{co}$. After that,
every time $\dynobj{i}{co}$ is freed, we non-deterministically set a
special variable $\mathit{fr}$ to \emph{true}. This allows us to
generate an assertion $p \neq \mathit{fr}$ for each location
containing \textsf{*p} or \textsf{free(p)}~\cite{MHSV18}.

\paragraph{\bf Memory Leaks}

To find a memory leak, we check whether, at the end of the program, there is an
object $\dynobj{i}{co}$ such that $\mathit{fr} \neq \&\dynobj{i}{co}$. If such
$\dynobj{i}{co}$ exists, there is a memory leak present (some object
represented by the corresponding $\dynobj{i}{}$ has has not been freed). On the
other hand, absence from memory leaks can only be proven for
programs without loops (or with loops that can be fully unwound). This is
because checking that $\mathit{fr}$ may be equal to a materialized object
$\dynobj{i}{co}$ is not sufficient to prove that all objects represented by the
corresponding abstract object $\dynobj{i}{}$ were freed~\cite{MHSV18}.

\section{Strengths and Weaknesses}


This year's improvements mostly influenced results in the MemSafety category
where 2LS narrowly missed the podium in 4th place. There were many new
benchmarks and a new sub-category (MemCleanup) whose benchmarks were
handled well by 2LS.


One of the main strengths of 2LS is verification of programs requiring
joint reasoning about shape and content of dynamic data structures.
There were no such benchmarks in previous SV-COMP editions, thus, we
contributed 10 of our own benchmarks.  Combining our shape domain with
the zones domain for value analysis allows 2LS to successfully verify
9 out of 10 of these benchmarks (the last one has timed out). None of
the other tools was able to verify more than 3 of these benchmarks.

Still, there remain a lot of challenges and limitations. The heap domain is
quite simple and over-approximates the heap too much to allow us to analyse
complicated properties of dynamic data structures. Moreover, reasoning about
array contents is still lacking, and the 2LS' algorithm $k$I$k$I does not yet
support recursion. Moreover, there is a large number of unconfirmed witnesses,
especially in the termination analysis (500 points lost).

\section{Tool Setup}

The competition submission is based on 2LS version 0.7.%
\footnote{%
Executable available at 
\textsf{https://gitlab.com/sosy-lab/sv-comp/archives-2020}}
The archive contains the binaries needed to run 2LS
(\textsf{2ls-binary, goto-cc}), and so no further installation is needed.
There is also a wrapper script \textsf{2ls} which is used by Benchexec
to run the tools over the verification benchmarks.  See the wrapper
script also for the relevant command line options given to 2LS.  The
further information about the contents of the archive could be find in the
\textsf{README} file.  The tool info module for 2LS is called
\textsf{two\_ls.py} and the benchmark definition file
\textsf{2ls.xml}.
As a back end, the competition submission of 2LS uses Glucose~4.0.
2LS competes in all categories except Concurrency and Java.

\enlargethispage{6mm}

\section{Software Project}

2LS is maintained by Peter Schrammel with pull requests contributed
by the community.%
\footnote{\textsf{https://github.com/diffblue/2ls/graphs/contributors}}
It is publicly available under a BSD-style license.
The source code is available at \textsf{http://www.github.com/diffblue/2ls}.

\bibliographystyle{splncs03}
\bibliography{bibliography}

\begin{thebibliography}{1}
\providecommand{\url}[1]{\texttt{#1}}
\providecommand{\urlprefix}{URL }

\bibitem{BJKS15}
Brain, M., Joshi, S., Kroening, D., Schrammel, P.: {Safety Verification and
  Refutation by $k$-Invariants and $k$-Induction}. In: SAS. LNCS, vol. 9291,
  pp. 145--161. Springer (2015)

\bibitem{CDK+17}
Chen, H.Y., David, C., Kroening, D., Schrammel, P., Wachter, B.: {Bit-Precise
  Procedure-Modular Termination Proofs}. TOPLAS  40 (2017)

\bibitem{DBLP:conf/tacas/ClarkeKL04}
Clarke, E.M., Kroening, D., Lerda, F.: A tool for checking {ANSI-C} programs.
  In: TACAS. LNCS, vol. 2988, pp. 168--176. Springer (2004)

\bibitem{MHSV18}
Mal{\'\i}k, V., Hruska, M., Schrammel, P., Vojnar, T.: Template-based
  verification of heap-manipulating programs. In: FMCAD. pp. 103--111 (2018)

\bibitem{MMSSVW18}
Mal{\'i}k, V., Marti{\v{c}}ek, {\v{S}}., Schrammel, P., Srivas, M., Vojnar, T.,
  Wahlang, J.: {2LS: Memory Safety and Non-termination (Competition
  Contribution)}. In: TACAS. pp. 417--421. Springer (2018)

\bibitem{SK16}
Schrammel, P., Kroening, D.: {2LS for Program Analysis (Competition
  Contribution)}. In: TACAS. LNCS, vol. 9636, pp. 905--907. Springer (2016)

\end{thebibliography}

\end{document}